\documentstyle[prl,aps,epsf]{revtex}
\draft
\begin{document}
\twocolumn[\hsize\textwidth\columnwidth\hsize\csname @twocolumnfalse\endcsname

\title{Structure and Melting of Two-Species Charged Clusters in a Parabolic Trap
} 
\author{J.A. Drocco$^{1}$, C.J. Olson Reichhardt$^{2}$, C. Reichhardt$^{2}$, 
and B. Jank{\' o}$^{1}$}
\address{ 
$^{1}$ Department of Physics, University of Notre Dame, Notre Dame,
Indiana 46556\\
$^{2}$ Theoretical Division, 
Los Alamos National Laboratory, Los Alamos, New Mexico 87545}

\date{\today}
\maketitle
\begin{abstract}
We consider a system of charged particles interacting with an
unscreened Coulomb repulsion in a two-dimensional
parabolic confining trap. 
The static charge on
a portion of the particles is twice as large 
as the charge on the remaining particles.  The particles separate into a shell
structure with those of greater charge situated farther from
the center of the trap.  As we vary the ratio of the number of particles of
the two species, we find that for certain configurations,
the symmetry of the arrangement of 
the inner cluster of singly-charged particles matches the 
symmetry of the outer ring of doubly-charged particles.  
These matching configurations have a higher melting 
temperature and a higher thermal threshold
for intershell rotation between the species
than the nonmatching configurations. 
\end{abstract}

\vskip2pc]
\narrowtext

Clusters of repulsive particles in confined traps 
have attracted considerable recent attention due to 
their applicability to a
wide variety of systems.  For example, 
two-dimensional (2D) clusters can represent electrons in quantum dots
\cite{Reed}
or on the surface of liquid helium \cite{Leiderer}, vortices 
in superfluids \cite{Kondo}, 
colloidal particles in circular traps \cite{Bubeck},
confined ferromagnetic particles \cite{Davidov},
and charged dust particles in plasma traps \cite{Jaun}. 
The 2D charged clusters
also resemble the problem of charge distribution studied
by Thomson in his ``plum-pudding'' model of the atom \cite{Thomson}.  

When confined to a parabolic trap, charged particles 
form a structure of concentric
rings, with the inner particles forming a distorted triangular lattice 
resembling a defected Wigner crystal, and the outer rings taking on
a more circular shape 
that conforms to the radial symmetry of the trap 
\cite{Bedanov94,Bedanov}.  Among the possible charge 
configurations are several ``magic'' arrangements
in which the number of particles is such that the shells 
form with only a few symmetrically distributed dislocations,
and so have a reduced total energy compared to what is predicted 
based on a semi-empirical approximation \cite{Koulakov}.
Bubeck et al. \cite{Bubeck}
have observed that certain colloidal clusters confined by a
circular hard-wall trap exhibit {\it two-stage} melting, in which 
intershell rotation between the outer two shells 
occurs at temperatures below the temperature at which
particles are able to jump between the shells.

Several explanations of this two-stage melting 
phenomenon have been proposed 
\cite{Bedanov94,Schweigert,Schweigert00,Lai}, 
all of which focus on the intershell 
rotation which occurs 
prior to the exchange of particles between shells.  Most plausible among
these is the theory that the rotation is due to an incommensuration 
between the shapes of the potentials created by the adjacent shells.
For this intershell rotation to occur, the inner shell configuration must
be sufficiently stable to have a melting temperature
higher than the threshold for intershell rotation.

In our simulation, we extend the confined charge system to include 
particles with two
distinct values of charge.  We find that the two
species separate into shells, with those of greater charge
located farther from the center of the trap.  This occurs due to the
fact that the
pinning force couples only to the position of the particle,
while the interparticle repulsion is a function of 
both position and charge.  Thus particles of greater charge are pushed 
farther up the walls of the parabolic trap.  If two-stage melting 
and intershell rotation is caused by an incommensuration
between the potentials formed by the particles, then 
we should be able to predict the occurrence of intershell rotation
based on the 
ratio of particles in the outer
shell of singly-charged particles to the number 
of doubly-charged particles.

We consider a system of $N_s + N_d$ charged particles interacting via an 
unscreened $1/r$
Coulomb repulsion, where $N_s$ is the number of single-charge particles 
with $q_s=1$ and
$N_d$ is the number of doubly-charged particles
with $q_d=2$.  The particles are free to 
move in two-dimensions 
but are confined by a parabolic trap centered at the origin and 
increasing radially as $r^{2}$.  The dimensionless Hamiltonian \cite{Bedanov}
for this system is:
\begin{equation}
H=\sum_{i=1}^{N_s+N_d} \sum_{j=i+1}^{N_s+N_d} 
\frac{q_{i}q_{j}}{|\vec r_{i}-\vec r_{j}|}+
A(\sum_{i=1}^{N_s+N_d}|\vec r_i|^{2})
\end{equation}
where $q_i$ ($\vec r_{i}$) is the charge (position) of particle $i$ 
and we fix the strength of the parabolic trap to $A=10$.
Charged colloidal particles in a confinement potential, 
such as those employed by Wei et al. \cite{Wei}, bear the closest 
resemblance to the Coulomb interaction used in this study.

Using a molecular dynamics (MD) simulation method, we initialize 
the system at high 
temperature, simulated by random Langevin kicks, and then slowly 
anneal it to a $T=0$ ground state
configuration.  We checked the 

\begin{figure}
\centering
\epsfxsize=3.5in
\epsfbox{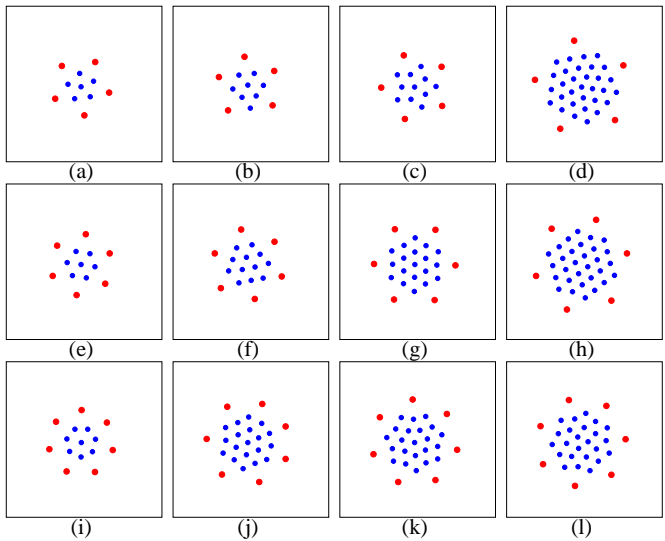}
\caption{Ground-state colloid configurations for: 
(a-d) $N_d=5$, (e-h) $N_d=6$, and (i-l) $N_d=7$, with $N_s =$
(a) 6, (b) 8, (c) 11, (d) 31, (e) 7, (f) 12, (g) 19,
(h) 30, (i) 8, (j) 19, (k) 22, (l) 24.  Small dots are the
singly charged particles and large dots are the doubly
charged particles.
}
\label{fig:catalog}
\end{figure}

\noindent
accuracy of this method by reproducing
 the ground state configurations for single-species
particle clusters found by Kong et al.\cite{Kong}, 
albeit with different confinement strength. 
Once we have
obtained the ground states of the two-species 
clusters, we slowly increase the temperature and observe the 
melting of the system.

Fig. ~\ref{fig:catalog} shows several
examples of the ground-state configurations 
we obtained for $N_d=5$, 6, and 7, respectively.  
For $N_d$=5, configurations with
$N_s=6$, 
11, and 31 have five-fold symmetry matching the number of doubly-charged
particles in the outer shell.  $N_s=8$ approximates a six-fold symmetric 
arrangement by substituting a singly-charged particle to fill a potential
well in the outermost shell.
For $N_d=6$, configurations with $N_s=7$ and $N_s=19$ have six-fold
symmetry.  Configurations with $N_s=12$ and $N_s=27$ have three-fold symmetry
and configuration $N_s=30$ 
has two-fold symmetry, which also are commensurate with
the arrangement of particles in the outer shell.
For $N_d=7$, $N_s=8$ and 22 are the only seven-fold symmetric inner particle
configurations. $N_s=23$ 
and 24 are nearly symmetric, however, with seven-fold symmetry in
the outer two shells of the inner particle configuration.
In general we find highly ordered structures when the 
smaller particles form a commensurate structure with the outer particles.
All the configurations are shown at \cite{Olson}. 
The general expressions $N_s=kN_d$ and $N_s=kN_d+1$, 
with $k$ a small integer, predict some
configurations that have rotational symmetry 
through an angle $2\pi/N_d$, such as $N_d=5$, $N_s=6$, but
do not work for all values of $k$
due to the fact that the outer particles may distort from a 
uniformly spaced arrangement in order to better accommodate the 
inner particles.
Although in this work we only present results 

\begin{figure}
\epsfxsize=3.5in
\epsfbox{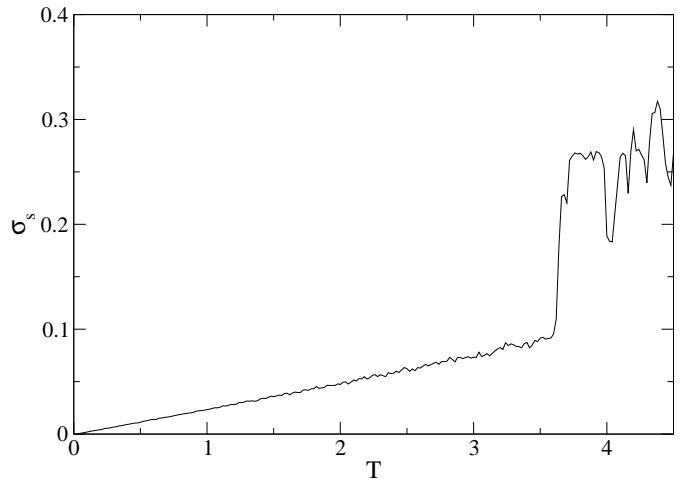}
\caption{Plot of $\sigma_s$ 
vs. temperature for 
$N_d=6$, $N_s=7$, averaged over five realizations.}
\label{fig:67radial}
\end{figure}

\noindent
for 
$q_d=2q_s$, we have considered other charge ratios, and
find that the more highly charged particles always move to the outside
of the trap,
and the same general commensurate-incommensurate phases occur. 
We have also performed simulations with larger $N_d>7$; 
however, we do not treat this case here since the outer particles begin 
to form multiple rings which have their own 
commensurate-incommensurate transitions.  In addition, we have run a 
set of simulations for $N_s<15$ 
with the same trap but with a $1/r^3$ interparticle
interaction potential, which corresponds to the interaction between
magnetic colloids.  We found that, in general, the ground states and
also the dynamical properties were qualitatively 
identical to those in the case of Coulomb repulsion for small $N_s$
 $(N_s<15)$.

\begin{figure}
\epsfxsize=3.5in
\epsfbox{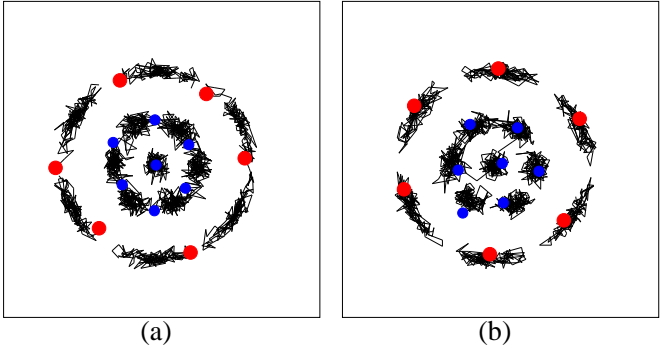}
\caption{Particle trajectories (lines) for $N_d=6$, $N_s=7$ 
at temperatures (a) $T=3.5$ with no intershell exchange
and (b) $T=3.6$ with intershell exchange.
}
\label{fig:6-7exchange}
\end{figure}

In order to confirm whether the highly ordered commensurate phases
are more stable we consider the melting of the two-species system.
We determine the temperature $T_e$
of the first exchange of particles between shells 
by measuring intershell exchange of the singly charged particles,
\begin{equation}
\sigma_s=\frac{1}{N_s}\sum_{i=1}^{N_s} {|\vec r_{i}(t)-\vec r_{i}(0)|}
\end{equation}
This gives the mean radial distance of the inner singly-charged
particles from their initial $T=0$ positions.
$\sigma_s$ shows a marked increase at $T_e$ when
the inner particles begin to jump between shells, as illustrated
in Fig.~\ref{fig:67radial} for a system with $N_s=7$ and 
$N_d=6$.
This configuration is highly stable, as shown in Fig.~\ref{fig:catalog}(e), 
with one 
central particle surrounded by an inner hexagonal shell of six 
singly-charged
particles and an outer hexagonal shell of six doubly-charged
particles. At $T=3.55$, $\sigma_s$ jumps
when one of the first-shell
particles exchanges with the central particle, as shown in the trajectories of 
Fig.~\ref{fig:6-7exchange}.

A similar measure, $\sigma_d$, tracks the doubly-charged particles;
however, as they never formed more than one shell

\noindent
for the parameters considered here, there were no intershell
exchanges and $\sigma_d$ only increased linearly with temperature.  
We observe exchanges of a singly-charged particle with 
a doubly-charged particle only at very high temperatures $T\gg T_e$, 
and these exchanges 
occur only for highly asymmetric and disordered configurations.

To track the onset temperature $T_r$ of intershell rotation between 
the two species when it occurs,
we use a second measure, $\Delta_\theta$:
\begin{equation}
\Delta_\theta=|\frac{1}{N_s}\sum_{i=1}^{N_s} (\theta_{i}(t)-\theta_{i}(0))-
\frac{1}{N_d}\sum_{j=1}^{N_d} (\theta_{j}(t)-\theta_{j}(0))| \ ,
\end{equation}
which gives the difference between the mean angular displacements of the two
species from their initial configurations.  
$\Delta_\theta$ increases when the shells 
slip past each other, but is insensitive to coherent rotation of the 
two species.  $\Delta_\theta$ becomes meaningless if the particles do not
maintain the same orientation with respect to the other particles of the 
same species, so it can detect intershell rotation only when this occurs 
{\it before} the onset of intershell exchange.  
$\Delta_\theta$ also detects relative slip between shells of the same 
particle species, which does not interest us here.  
Such same-species shell slips were generally limited to the 
erratic rotation of two particles at the center of the configuration, 
and produced a sufficiently small change
in $\Delta_\theta$ to be distinguished easily 
from a genuine rotation relative to the other species.  
For configurations with a single
central particle located roughly at the origin, the angular 
displacement of that particle was excluded
from $\Delta_\theta$ in order to reduce noise.

\begin{figure}
\epsfxsize=3.5in
\epsfbox{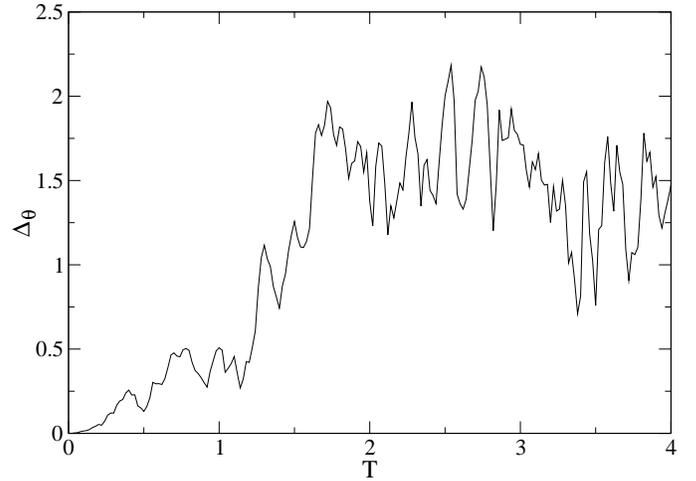}
\caption{Plot of $\Delta_\theta$ vs. temperature for 
$N_d=7$, $N_s=7$, averaged over five realizations.
}
\label{fig:7-7ang}
\end{figure}

An example of intershell rotation measured by $\Delta_\theta$ is shown
in Fig. ~\ref{fig:7-7ang}, where we plot 
$\Delta_\theta$ vs temperature
for a configuration of $N_s=7$ and $N_d=7$. 
In this case, the configurations of each of the 
two species are highly stable independently,
with the singly-charged particles forming a hexagonal ring around a central
particle.  However, there is an incoherence between the hexagonal ring of
singly-charged particles and the seven-particle ring of 

\begin{figure}
\epsfxsize=3.5in
\epsfbox{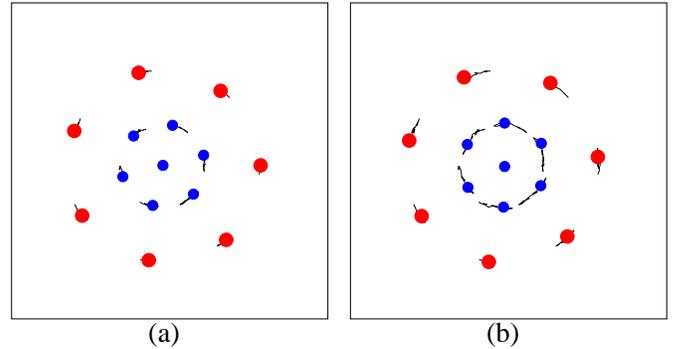}
\caption{Particle trajectories over a short time interval
for $N_d=7$, $N_s=7$ at (a) $T=0.6$ with no intershell rotation and 
(b) $T=0.7$ with intershell rotation.
}
\label{fig:7-7rotate}
\end{figure}

\noindent
doubly charged
particles which surrounds it.
A substantial increase in ${\Delta_\theta}$ occurs at
$T\approx 0.65$, corresponding to a slipping between the outer shell
and the inner cluster of particles.
The trajectory plot, Fig. ~\ref{fig:7-7rotate}, also 
indicates this slipping, as the 
trails of the inner shell particles become larger than those of the
outer shell particles at this temperature.  
Since the onset of intershell rotation at $T_r$ is more 
gradual than the intershell exchange transition,
we define $T_r$ to occur when $\Delta_\theta$ exceeds
a threshold value of $\Delta_\theta=\frac{\pi}{N_d}$, corresponding to 
an angular
displacement of half of the angular distance between neighboring particles
in the outer shell of singly-charged particles. 

\begin{figure}
\centering
\epsfxsize=3.5in
\epsfbox{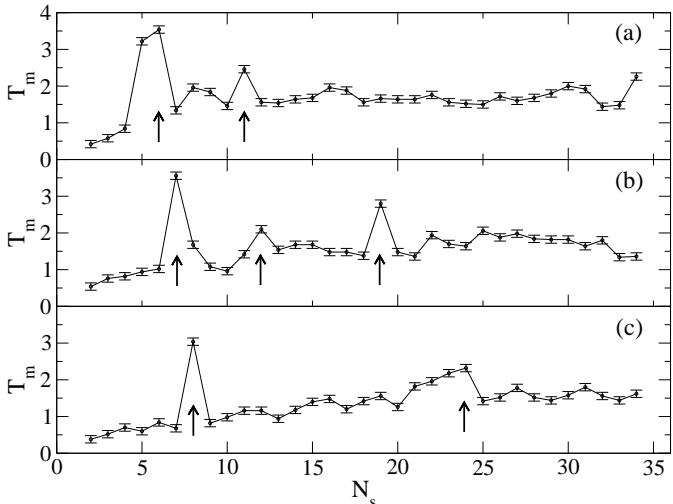}
\caption{Melting temperature $T_m$ vs $N_s$ for 
(a) $N_d=5$, (b) $N_d=6$, and (c) $N_d=7$. 
}
\label{fig:tempcurve}
\end{figure}

The matching of the inner particle symmetry with the outer shell
symmetry produces an elevated melting threshold for
commensurate 
configurations.  The melting temperature $T_m$ is taken to be the lower
of $T_e$ or $T_r$.
In Fig.~\ref{fig:tempcurve}(b) we plot $T_m$
for a range of $N_s$ at fixed $N_d=6$.  
Figs.~\ref{fig:tempcurve}(a)
and (c) show $T_m$ for $N_d=5$ and $N_d=7$, respectively.
For each configuration, we averaged $\sigma_s$ and $\Delta_\theta$ over 
five realizations to reduce error.
In Fig.~\ref{fig:tempcurve}(b), $T_m$ for $N_d=6$ has a peak at $N_s=19$, 
another perfect triangular arrangement. 
We also find an 

\noindent
elevated melting temperature for 
$N_d=12$, which has threefold symmetry.
As seen in Fig.~\ref{fig:tempcurve}(a),
configuration $N_d=5$ and $N_s=11$ also exhibits a higher than
average melting temperature.  This is expected since the configuration 
$N_s=11$
has five-fold symmetry, commensurate with the number of outer particles.
Fig.~\ref{fig:tempcurve}(c) shows high $T_m$ values for $N_d=7$ at
$N_s=23$ and $N_s=24$,
which is 
expected due to the matching seven-fold symmetry of 
the outer three rings.  The 
inner part of the configuration $N_d=7$, $N_s=19$ forms the same highly stable 
perfectly hexagonal arrangement of $N_d=6$, $N_s=19$, 
and thus has a high threshold
for intershell exchange between the singly-charged inner particles.
However, the incommensuration with the 7 particles on the outside yields
a low threshold for intershell rotation.

There is an obvious shift in the peaks of Fig.~\ref{fig:tempcurve}(a-c) as the 
number of outer-shell doubly-charged particles changes.  Most notably,
the highest melting temperature for each $N_d$ occurs for 
$N_s=N_d+1$, when the singly-charged particles form a central particle
surrounded by a shell that is 
commensurate with the outer particles.
As expected, the highest melting temperature  
occurs for $N_d=6$ and $N_s=7$, as this arrangement matches the 
triangular lattice of the Wigner crystal.  However, melting
temperatures nearly as high occur for $N_d=5$ and $N_s=5$ and 6.
We believe that the smaller size of the crystal 
in this case exaggerates the effect
of the commensuration between rings, despite the dissimilarity with
the bulk lattice configuration. 

In conclusion 
we have investigated the structure and melting of 
two species of charged particles in a parabolic trap.
The more highly charged particles form an outer ring.
Highly ordered clusters occur when the structure of the 
central cluster of singly charged particles
is commensurate with the outer ring of doubly charged particles. 
We observe variations in the melting temperatures of the two species
clusters, with
elevated melting thresholds for intershell rotation
and intershell particle exchange when a commensuration 
occurs between the symmetry of the inner cluster and the outer ring.

This work was supported by the U.S. Department of Energy
under Contract No. W-7405-ENG-36 and by the U.S. DoE, Office
of Science, under Contract No. W-31-109-ENG-38. BJ and JD were supported by
NSF-NIRT award DMR02-10519 and the Alfred P. Sloan Foundation.

\end{document}